\documentclass[aps,prl, twocolumn, superscriptaddress,groupedaddress]{revtex4}   
\usepackage{graphicx}   
\usepackage{dcolumn} 
\usepackage{bm}       
\usepackage{amssymb}   
\usepackage{amsmath}
\usepackage{ulem} 
\begin{document} 
\title{Environmental Disorder Regulation of Invasion and Genetic Loss}
\author{Youness Azimzade}
\email[Email:~]{ y_azimzade@ut.ac.ir} 
\affiliation{Department of Physics, University of Tehran, Tehran 14395-547, Iran}
\affiliation{Mathematical Oncology Laboratory, Universidad de Castilla-La Mancha,  13071 Ciudad Real, Spain} 
\author{Mahdi Sasar} 
\affiliation{Department of Physics, University of Tehran, Tehran 14395-547, Iran}
\author{V\'{\i}ctor M. P\'erez Garc\'{\i}a} 
\affiliation{Mathematical Oncology Laboratory, Universidad de Castilla-La Mancha,  13071 Ciudad Real, Spain} 
\date{\today}
\begin{abstract}
	Many physical and natural systems, including the population of species, evolve in habitats with spatial stochastic variations of the individuals motility. We study here the effect of those fluctuations on invasion and genetic loss.
A Langevin equation for the \textit{position} and \textit{border} of the invasion front is obtained. A striking result is that small/large fluctuations of diffusivity suppress/intensify genetic loss. Our findings reveal the potential role of environmental fluctuations as a regulating factor for genetic loss and provides a simple explanation for the regional differences on the intensity of genetic drift observed during the final stages of human evolution and in tumor mutational landscapes.
\end{abstract}
\maketitle 

\textit{\textbf{Introduction}} The spread of populations is a phenomenon bearing resemblances with other physical diffusion processes, such as those ruled by reaction-diffusion equations  \cite{kondo2010reaction}. 
Migrations, invasions of different populations and even tumor growth are described in a first approximation by simple mathematical models such as the Fisher-Kolomogorov-Petrovsky-Piskunov (FKPP) equation \cite{murray2003mathematical}. 
Similar models have been used recently to describe genetic drift \cite{hallatschek2011noisy, hallatschek2009fisher}.  

Substantial physical and mathematical work has contributed to the understanding of processes ruled by deterministic reaction-difusion dynamics, however it has been recently pointed out that fluctuations may have non-trivial effects on the invasion dynamics \cite{birzu2018fluctuations}. 

The FKPP has been used to describe the dynamics of infiltrative tumors throughly in the last 20 years \cite{murray2003mathematical, swanson2003virtual, mandonnet2003continuous}.  
The physical properties of the tumor microenvironment  such as  host tissue stiffness \cite{bordeleau2017matrix, mason2013tuning}   have an effect on cells.  These properties,  including host tissue stiffness, exhibit spatial fluctuations \cite{jamin2015exploring, plodinec2012nanomechanical} what could influence cellular invasion processes. Similarly, for entities moving within a habitat, the ability to move may depend on space due to `random' variations in the physical properties of the environment \cite{howell2018increasing}.
As a result, studying invasion in those environments requires the study of mathematical models with spatially fluctuating diffusion constant. Many other biological systems exhibit similar heterogeneities. 

An interesting process associated with invasion is genetic drift in which the frequency of different alleles changes due to random fluctuations and may lead to the extinction/fixation of some of them \cite{allendorf1986genetic}.  During invasion, genetic drift   plays a central role in  population dynamics \cite{hallatschek2007genetic, slatkin2012serial, birzu2019genetic}. As such, genetic drift analysis has been under theoretical and experimental investigation for different types of expanding populations \cite{hallatschek2007genetic, reiter2014range}.  One of the most active fields in this area concerns human genetics. After experiencing a bottleneck between 100,000 and 60,000 years ago, modern humans started to expand out of Africa with a velocity close to 1 km/year and lost genetic diversity during that expansion \cite{cavalli1993demic}. 
However, the intensity of genetic drift is not the same for all areas and measurement of the human data has shown larger genetic loss in East Asians than in Europeans \cite{keinan2007measurement}, but the reason behind these differences remains unclear. 

In this letter, we study the effect of spatial fluctuations of the diffusion constant on invasion front wanderings and genetic loss. First, we perform a perturbation analysis of the proposed equation to obtain the behavior of  the front position and confirm the findings through numerical simulations.  Then we show, both numerically and analytically, how these fluctuations affect the population composition during invasion and thus play a role in the regulation of genetic loss within invading populations. Our findings may provide further insight on the genetic loss observed in ancient human genetic data and on the problem of tumor heterogeneity.

\begin{figure} [!htb]
	\includegraphics[width=0.49\linewidth]{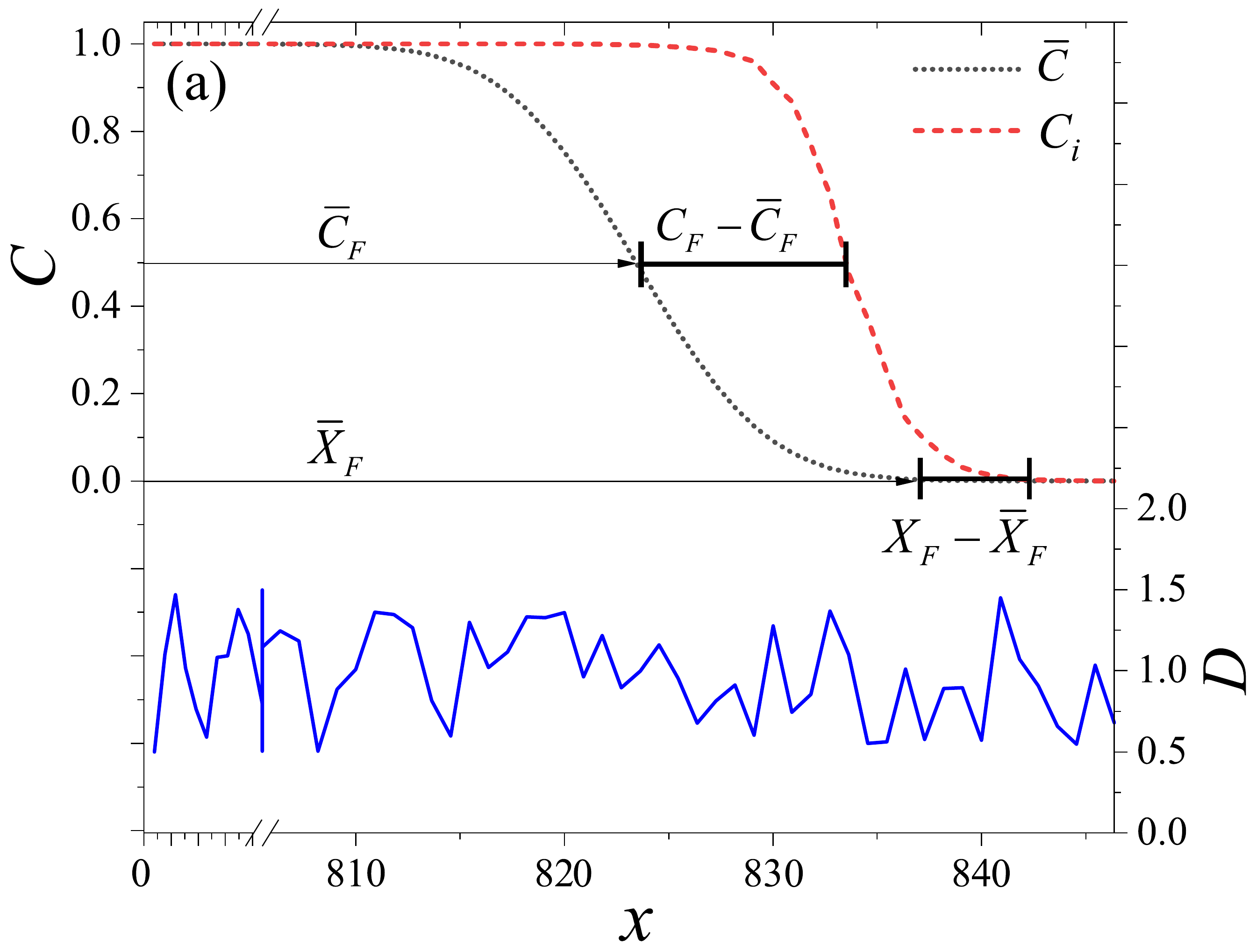}
	\includegraphics[width=0.49\linewidth]{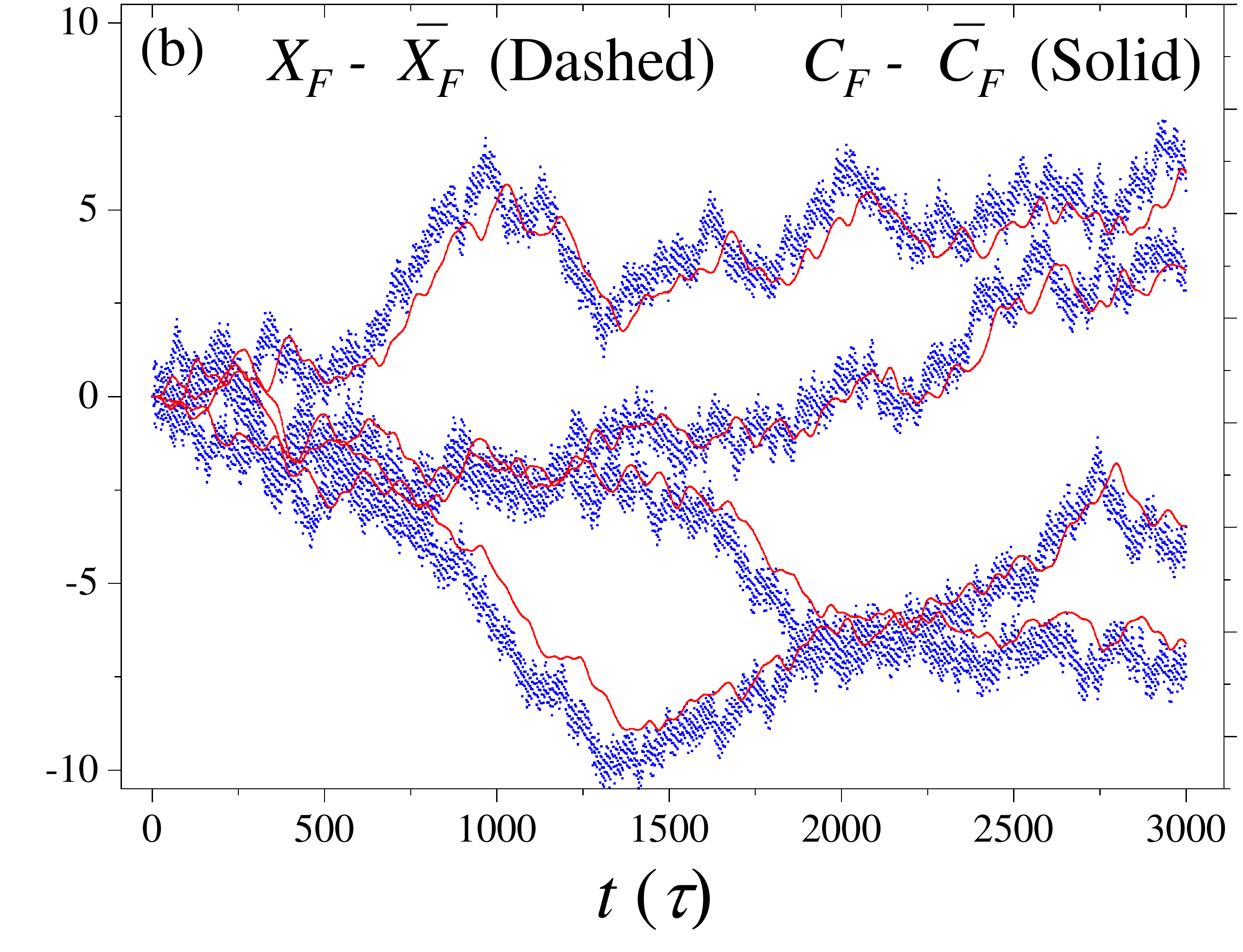}  
	\caption{(a) Schematic illustration of the model and parameters. Shown are a realization of $D$ for $\xi =0.5$ (lower blue curve) and the ensemble average solution (black dotted line) versus a realization of the stochastic  FKPP for parameter values of $\xi=0.5$, $R=0.01$ and $D_0=0.25$. The differences between the ensemble average values $\bar{C}_F$ and $\bar{X}_F$ and the realization values ${C}_F$ and ${X}_F$ are also indicated with a solid black line. (b) Four realizations of $X_F-\bar{X}_F$ and $C_{F}-\bar{C}_F$ for $R=0.01$ and $\xi=0.5$ and $D_0=0.25$.} 
	\label{FIG1}
\end{figure}

\textbf{\textit{Model}.} The dynamics of invasion processes will be described in this paper using the equation
\begin{equation}
\frac{\partial C}{\partial t}=RC(1-C)+\nabla(\bar{D}\nabla C),
\label{eq:3}
\end{equation}
where $C(x,t)$ is the number of individuals in units of the system carrying capacity, $R$ is the growth rate, 
$\bar{D}=D_0(1+\xi f(\boldsymbol{x}))$ is the diffusion constant, and $f$ is an uniform white noise in the range $[-1,1]$. Thus, our model is a 
FKPP equation with a spatially random diffusion.
We will study the motion of the invasion interface using the {\it front position}, defined as $C_F=\int_{0}^{\infty}C(x,t)dx$ and the {\it border location} $X_F$, defined as the point where $C(X_F,t)=10^{-10}$. Both quantities are shown in Fig. \ref{FIG1}.

Numerical simulations show that spatial fluctuations in $D$ lead to a deviation from deterministic behavior. This deviation is associated with fluctuations of both the front position and border.  Figure \ref{FIG1}(b) shows some examples of the fluctuating front dynamics. 

\textit{\textbf{Invasion Front Wanderings}.} 
In this paper, we will consider (\ref{eq:3}) in one spatial dimension
\begin{equation}
\frac{\partial C}{\partial t}=\frac{\partial}{\partial x}\left[\left(D_0+\xi f(x)\right)\frac{\partial C}{\partial x}\right]+RC(1-C)
\label{eq:7}
\end{equation}
Where $f(x,t)$ is the perturbing function and $\xi$ is a small dimensionless parameter that controls the strength of the perturbation. 
Near the invasion front the cell density satisfies $C \ll 1$. Thus, some insight on the  front dynamics in Eq. (\ref{eq:3}) can be obtained by linearizing Eq. (\ref{eq:3}).  To construct aproximate solutions we proceed perturbatively \cite{mikhailov1983stochastic},  writing the density as
\begin{equation}
C(\zeta,t) \approx C_0(\zeta+\eta(t),t)+ \delta C_1(\zeta,t).
\label{Mikhailov}
\end{equation}
where $C$ is written in the comoving frame and $\zeta = x-vt$. 
Furthermore, $C_0$ is assumed to satisfy the linearized equation with $\xi=0$, i.e.
\begin{multline} 
\frac{\partial C_0(\zeta,t)}{\partial t}-\hat{\Gamma} C_0(\zeta)=\frac{\partial C_0(\zeta,t)}{\partial t}  \\ -\bigg(D_0 \frac{\partial^2}{\partial\zeta^2} + v\frac{\partial}{\partial\zeta} + R\bigg)C_0(\zeta,t) = 0.
\label{Comoving}
\end{multline}
Which has the solution
\begin{equation}
C_0(\zeta,t) = \frac{1}{\sqrt{4\pi D_0 t}}e^{-\frac{1}{2}\sqrt{\frac{R}{D_0}}\zeta}e^{-\frac{\zeta^2}{4D_0 t}}.
\label{Comoving3}
\end{equation}
The first term in Eq. (\ref{Mikhailov}) describes the effects of the perturbing function $f(x)$ on the position of the propagating front, while the second term contains the first-order changes in the front shape. To find the effective diffusion coefficient for the fluctuating front, it is sufficient to solve Eq. (\ref{eq:3}) using (\ref{Mikhailov}) for $\eta(t)$  \cite{mikhailov1983stochastic,birzu2018fluctuations}. 
Asymptotically  $(t \gg 1/R)$, thus $v$ can be assumed to be equal to $2\sqrt{R D_0}$ \cite{brunet2001effect}. In moving to co-moving reference frame, $f(x) $ becomes  $f(\zeta , t) $.   However, if  we consider   $f(x)$  to be smooth enough or $v$ to be small,  in a co-moving reference frame, we still can have temporally quenched fluctuations or  $f (\zeta , t)\sim f (\zeta )$ in perturbation range. 
Plugging (\ref{Mikhailov}) into  (\ref{eq:7}) we get \begin{equation}
\frac{\partial \delta C_1}{\partial t} - \hat{\Gamma}\delta C_1 + \dot{\eta}(t) C_0(\zeta,t) = \xi \frac{\partial}{\partial \zeta} \bigg(f(\zeta) \frac{\partial}{\partial \zeta}C_0(\zeta,t)\bigg).
\label{Comoving2}
\end{equation}
The operator $\hat{\Gamma}$ is not self-adjoint, its adjoint being $\hat{\Gamma^\dagger}=D_0 \dfrac{\partial^2}{\partial\zeta^2} - v\dfrac{\partial}{\partial\zeta} + R$.  Next, we multiply Eq. (\ref{Comoving2}) by the eigenfunction of $\hat{\Gamma^\dagger}$ with 0 eigenvalue (which is $e^{\sqrt{\frac{R}{D_0}}\zeta}$) and integrate over $\mathbb{R}$, to get     
\begin{equation}
\dot{\eta}(t) =\xi\dfrac{\int_{-\infty}^{\infty} e^{\sqrt{\frac{R}{D_0}}\zeta}\bigg(f(\zeta) C'_0(\zeta,t)\bigg)' d\zeta}{\int_{-\infty}^{\infty}e^{\sqrt{\frac{R}{D_0}}\zeta}C'_0(\zeta,t) d\zeta},
\label{FinalStep2}
\end{equation}
what leads to
\begin{equation}
\eta(t) =\xi\int_{0}^{t} d\tau e^{-\frac{R\tau}{4}}
\int_{-\infty}^{\infty} d\zeta e^{\sqrt{\frac{R}{D_0}}\zeta}f(\zeta) C'_0(\zeta,\tau).
\label{FinalStep4}
\end{equation}
Then, the effective diffusion coefficient would be \cite{birzu2018fluctuations}
\begin{multline}
D_{C} = \dfrac{\langle \eta^2(t) \rangle}{2t}= \frac{\xi^2}{2t}\int_{0}^{t}d{T_1}\int_{0}^{t}d{T_2}\int_{-\infty}^{\infty}  \\ 
C'_0(\zeta,T_1)C'_0(\zeta,T_2)      e^{-\frac{R T_1}{4}}e^{-\frac{R T_2}{4}}e^{2\sqrt{\frac{R}{D_0}}\zeta}  d\zeta.
\label{FinalStep6}
\end{multline}
Now, we  perform an ensemble average over $\eta^2(t)$ using the fact that $\langle f(x)f(y) \rangle = \delta(x-y)$ for a normal distribution of the noise term. Some insight can  be obtained from Eq. (\ref{FinalStep6}) if we use dimensionless parameters $\tau_i=\frac{T_i}{t}$ and $\sigma= \zeta\sqrt{R/D_0}$. In other words
\begin{multline}
D_{C} = \xi^2 \dfrac{\sqrt{R}}{32\pi D^{3/2}_0}\int_{0}^{1}d\tau_1\int_{0}^{1}d\tau_2\int_{-\infty}^{\infty}d\sigma  e^{-R t\frac{(\tau_1+\tau_2)}{4}} \\  \times \dfrac{\left(1+\dfrac{\sigma}{Rt\tau_1}\right)\left(1+\dfrac{\sigma}{Rt\tau_2}\right)}{\sqrt{\tau_1\tau_2}}e^{-\frac{\sigma^2}{4Rt\tau_1}}e^{-\frac{\sigma^2}{4Rt\tau_2}}e^{\sigma}.
\label{FinalStep7}
\end{multline}
Equation (\ref{FinalStep7}) gives the effective diffusion coefficient for the stochastic behavior of the front. For a diffusive behavior, we would expect this coefficient to become constant asymptotically.
Thus, we can approximate the integral as $t\rightarrow \infty$ 
to obtain
\begin{equation}\label{Difi}
D_{C}(t\rightarrow\infty) =\dfrac{1}{8} \xi^2 \sqrt{R D_0}.
\end{equation}
Thus, spatial fluctuations in the diffusion constant of amplitude $\xi$ act as a nonlinear regulating factor for the invasion front wanderings, the dependence of the diffusion constant being  proportional to $\xi^2$.   

To numerically validate this result, we  discretized  Eq. (\ref{eq:3}) using a finite difference method, what leads to  a  master equation  for the population density  at each point  \cite{sahimi1983stochastic}:  
\begin{equation}
dC= P^{\pm}[C(x)-C(x \pm \Delta)]   +R(C)C,
\label{eq:4}
\end{equation} 
where $P^{\pm }$  stands for density flow rate in negative (positive) $x$ direction. For an homogeneous environment, $P^{\pm}$   is related to the diffusion constant as: $ P^{\pm}=\tau D/\Delta^{2}$ in which $\Delta$  
and $\tau$ are spatial discretization length  and time step respectively. For heterogeneous environments, we set the fluctuation length to be equal to our discretization length, $\Delta$, and we have:
$P^{\pm}=\tau D_0(1+\xi w^{\pm})/2\Delta^{2} $. 
Finally, we considered the flow between neighboring units to be integer multiples of $10^{-10}$. 
\begin{figure} [!htb]
	\includegraphics[width=0.49\linewidth]{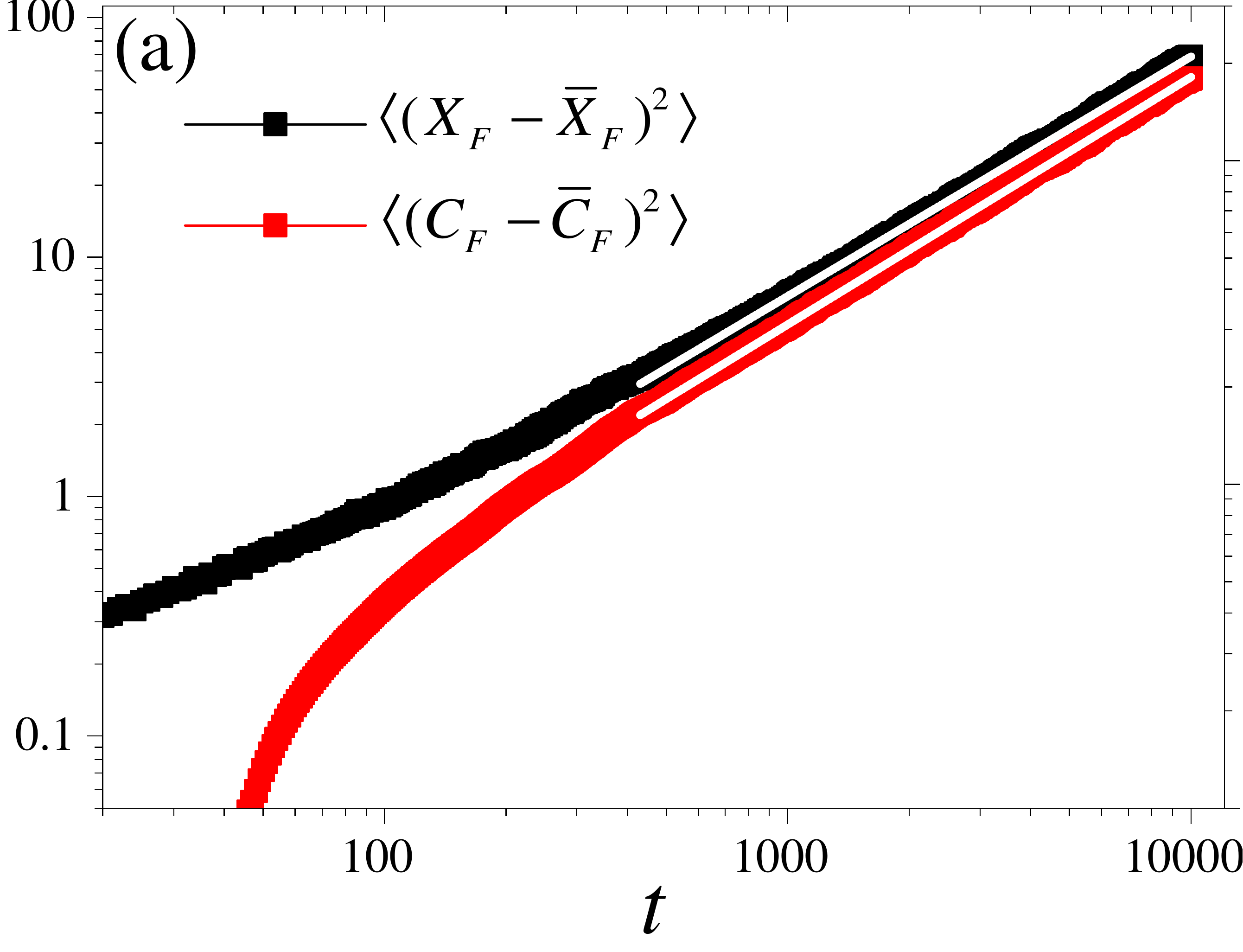}
	\includegraphics[width=0.49\linewidth]{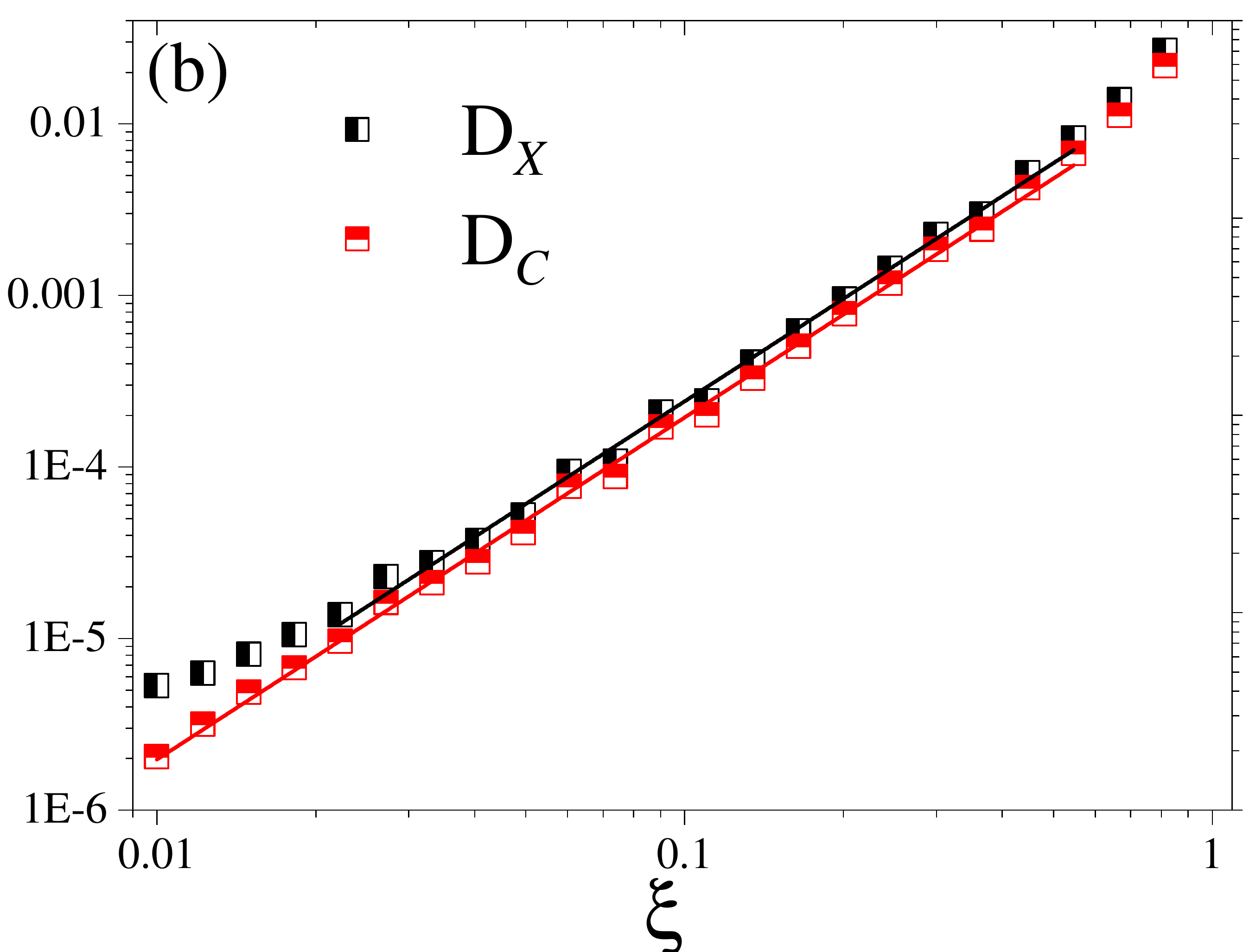}
	\includegraphics[width=0.49\linewidth]{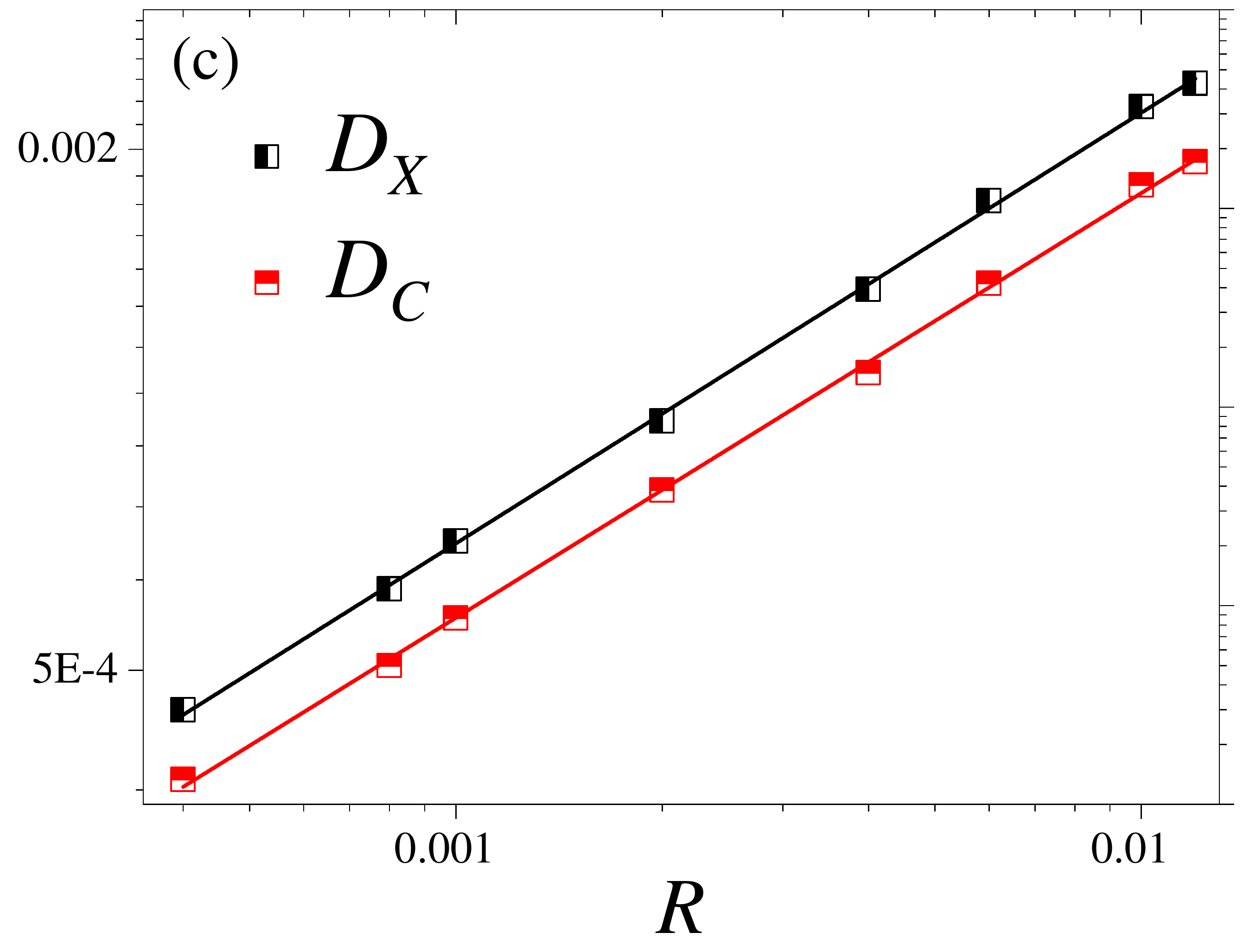}
	\includegraphics[width=0.49\linewidth]{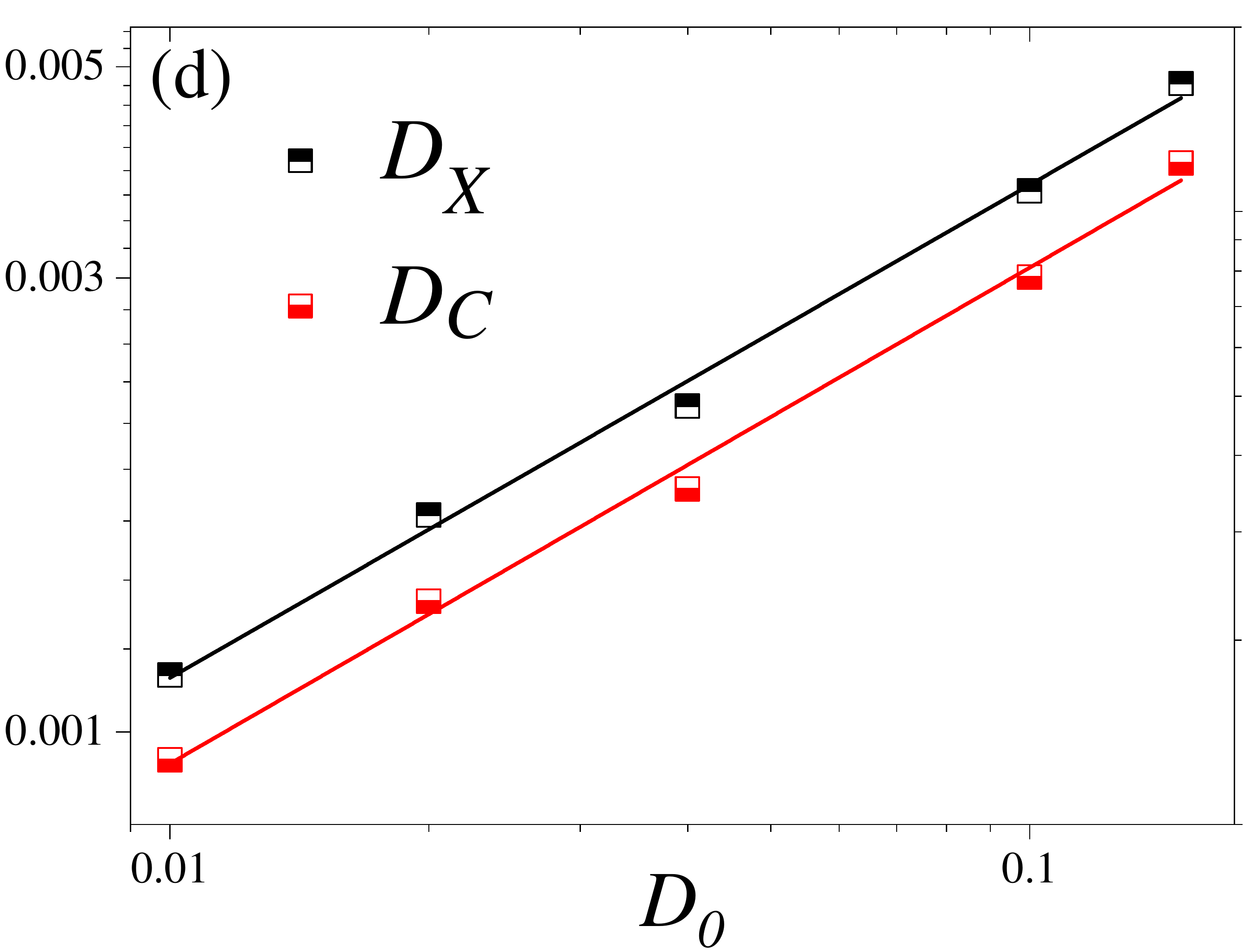} 
	\caption{(a)  $\langle(X-\bar{X})^{2}\rangle$ and $\langle(C_{F}-\bar{C}_F )^{2}\rangle$  versus time in ($\log / \log$ axes). The linear behavior and slope equal to one at large time scales guarantee a random walk like behavior and thus we can define a diffusion constant for each variable.  
		(b) $D_X$ and $D_C$ versus $\xi$ for $R=0.01$. The best fit obtaioned is $D_C\propto\xi^{2.00\pm0.01}$ for large range of $\xi$ values.  
		(c) Dependence of $D_X$ and $D_C$ on $R$. Slopes for $D_X$ and $D_C$ are $0.50\pm0.02$ and $0.5\pm0.02$ for $\xi=0.5$. ($R^2 = 0.996$).
		(d) Dependence of $D_X$ and $D_C$ on $D_0$ for $R=0.01$ and $\xi=0.5$. The best fit using least squares method is $D_X\propto D_0^{0.50\pm 0.02}$ and $D_C\propto D_0^{0.50\pm 0.02}$ ($R^2 = 0.995$), in agreement with Eq. (\ref{Difi}).}
	\label{FIG2}
\end{figure}

To study if the invasion border and front position obey a Langevin equation, we plotted $\langle(X_F-\bar{X}_F)^{2}\rangle$ and $\langle(C_{F}-\bar{C}_F)^{2}\rangle$ versus time.  As Fig. \ref{FIG2}(a)  shows, the log/log diagram of both quantities has slope one which means that we can obtain the corresponding diffusion constants, $D_X$ and $D_C$. We next studied numerically the dependence of both parameters on $\xi$, $R$ and $D_0$ and obtained  values in agreement with Eq. (\ref{Difi}) (see Fig. \ref{FIG2}).

\textit{\textbf{Genetic loss}.}  The effect of environmental factors, such as diffusion constant fluctuations in space on genetic loss has remained largely unexplored. In this part, we study the effect of fluctuations on genetic loss. Let us consider two mixed populations $C_1(x,t), C_2(x,t)$  which compete over space following the equation
\begin{equation}
\frac{\partial C_i}{\partial t} = \frac{\partial}{\partial x}\bigg[D_0 \left( 1+\xi f(x) \right) \frac{\partial C_i}{\partial x} \bigg]   + R C_i (1 - C_1 - C_2),
\label{AddedEq01}
\end{equation} 
for $i=1,2$, that is a two-population extension of Eq. (\ref{eq:3}). 
We will work in the regime of weak perturbations and assume that initial data for the two populations differ by a small  amount $2\epsilon$, i.e. $C_1(x,0) =0.50-\epsilon, C_2(x,0) =0.50+\epsilon$, with $\epsilon >0$. Let us define $C = C_1 + C_2$ and $\Upsilon = C_2 - C_1$. These quantities satisfy the equations \begin{subequations}
	\begin{eqnarray}
	\frac{\partial C}{\partial t} = \frac{\partial}{\partial x}\left[ D_0 \bigg( 1+\xi f(x) \bigg ) \frac{\partial C}{\partial x} \right] + R C(1 - C),     
	\\ 
	\frac{\partial \Upsilon}{\partial t} = \frac{\partial}{\partial x}\left[ D_0 \bigg( 1+\xi f(x) \bigg ) \frac{\partial \Upsilon}{\partial x} \right] + R \Upsilon (1 - C). 
	\label{AddedEq02}
	\end{eqnarray}
\end{subequations}
In the absence of fluctuations, if the two populations were to start from the same initial populations, $\Upsilon (x,t) = 0$. To gain insight into the effect of the noise in the dynamics of the two populations, we define $\kappa(t)$ as follows.  
\begin{eqnarray}
\kappa(t) = \int_{-\infty}^{\infty} dx \; \Upsilon^2(x,t).
\label{AddedEq03}
\end{eqnarray}
Differentiating $\kappa(t)$ with respect to $t$, yields: 
\begin{multline}
\dot{\kappa}(t) -2 R \kappa(t) = 2 R \int_{-\infty}^{\infty}dx \; C(x,t) \Upsilon^2(x,t) \\  
- 2 D_0 \int_{-\infty}^{\infty}dx \;\bigg( 1+\xi f(x) \bigg )\bigg(\frac{\partial\Upsilon}{\partial x}\bigg)^2.  
\label{AddedEq05}
\end{multline}
For two very close initial populations, for $x<x_{front}$ when $C \approx 1$ then $\Upsilon \approx 0$, and when the two populations diverge and $\Upsilon \approx 1$, then $C \approx 0$. Thus, we can neglect the first term in the right-hand side of Eq. (\ref{AddedEq05}).  
For small perturbations, $C_i \approx C^0_i + \xi C^1_i$, and thus, $\Upsilon \approx \Upsilon_0 + \xi \Upsilon_1$. Keeping the lowest order terms  in $\xi$, we get
\begin{equation}
\dot{\kappa}(t) -2 R \kappa(t) \approx -2 D_0 \int_{-\infty}^{\infty}dx \;\bigg( 1+\xi f(x) \bigg )\bigg(\frac{\partial\Upsilon_0}{\partial x}\bigg)^2,
\label{AddedEq07}
\end{equation}  
from where we can compute the fluctuations in $\kappa$.  
\begin{eqnarray}
\kappa_1(t) = -2\xi D_0 \int_{0}^{t}dt' \; G(t-t')\int_{-\infty}^{\infty}dx \; f(x)\bigg(\frac{\partial\Upsilon_0}{\partial x}\bigg)^2.
\label{AddedEq10}
\end{eqnarray}
where $G(t-t')$ is the green function for $\dot{\kappa}(t) -2 R \kappa(t) =0$. Finally, using that $\langle f(x)f(y) \rangle = \delta(x-y)$ we obtain
\begin{eqnarray}
\langle \Delta\kappa^2 \rangle = 4\xi^2 D^2_0\int_{0}^{t}\int_{0}^{t} dt'dt'' \;  G(t-t')G(t-t'') \nonumber \\  \int_{-\infty}^{\infty} dx' \; \bigg(\frac{\partial\Upsilon_0(x',t')}{\partial x'}
\frac{\partial\Upsilon_0(x',t'')}{\partial x'}\bigg)^2.
\label{AddedEq11}
\end{eqnarray}
When the two populations start from non-equal initial populations $\Upsilon \neq 0$, therefore, the integral in Eq. (\ref{AddedEq11}) is non-zero, and $\kappa$ fluctuates due to the noise.   

To gain more detailed understanding of effect  these fluctuations, we solved   Eq.  \eqref{AddedEq01} numerically.  For non-zero $ \epsilon$  in  an homogeneous environment we know that $C_1 $  will be extinct, due to its smaller initial value  (see Figure 3a, $C_{1}^{(\xi = 0)}$). Figure 3(a) shows that weak fluctuations delay the extinction process (see $C_{1}^{(\xi = 0.1)}$).  Interestingly, strong fluctuations have a substantially different effect.  As  Figure 3(a) shows, since they enhance the extinction process (see $C_{1}^{(\xi = 0.9)}$).   To further study the effect of  noise amplitude  $\xi$, on genetic loss, we studied the behavior of heterozygosity, defined as $H= \langle  \int  C_1 C_2  dx\rangle$ at the invasion front  where $ C_1+C_2 <1$.   
Heterozygosity, which quantifies coexistence of populations in invasion front,  gives us a  measure of genetic loss. As Figure 3(b) shows,  diffusion constant fluctuations decrease genetic loss for  $0 <\xi <0.2$ and increase it for $\xi >0.2$. 
\begin{figure} [!htb]
	\includegraphics[width=0.49\linewidth]{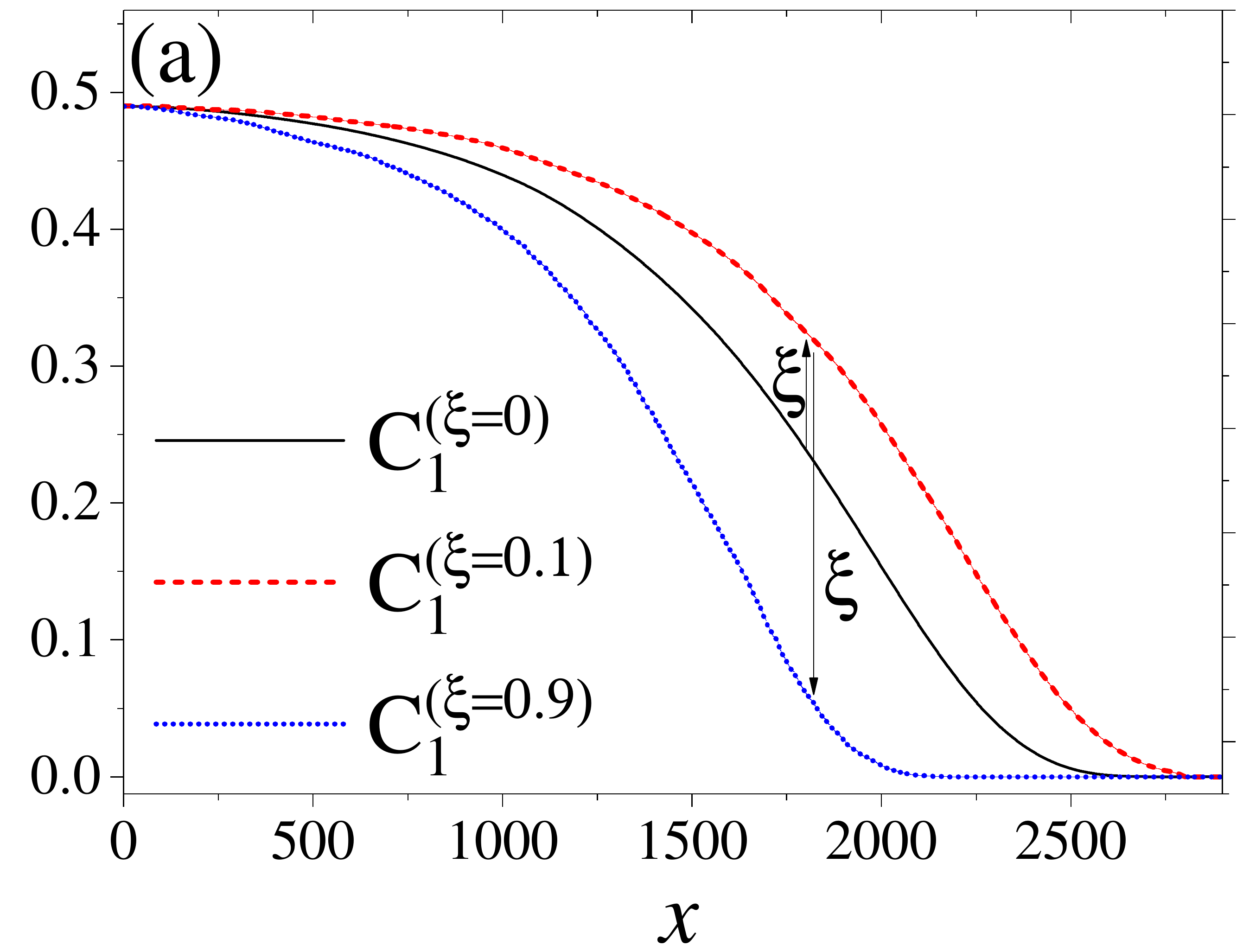}
	\includegraphics[width=0.49\linewidth]{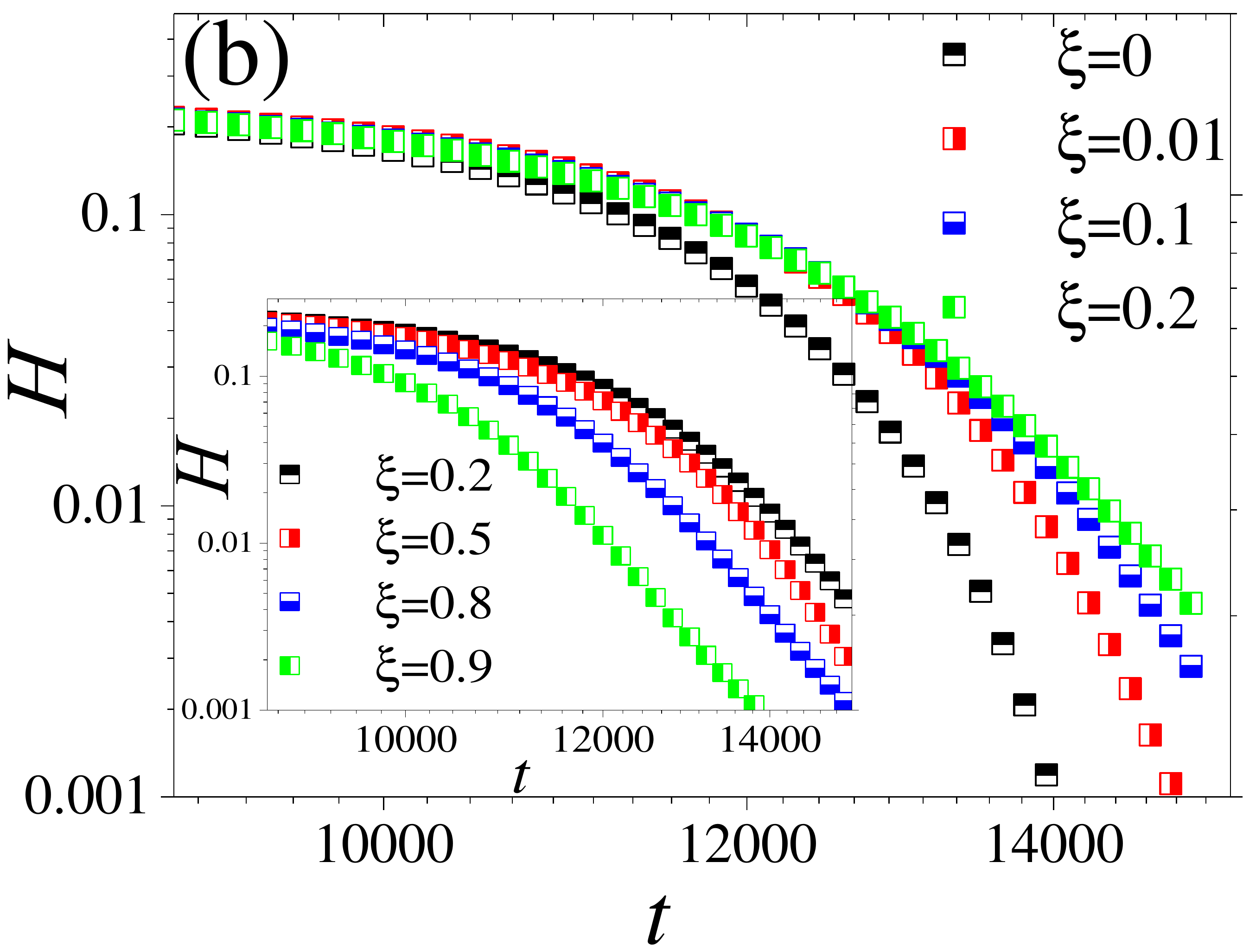}   
	\caption{(a) Plots of the densities $C_{1}^{(\xi = 0)}(x,t)$, $C_{1}^{(\xi  = 0.1)}(x,t)$, and $C_{1}^{(\xi = 0.9)}(x,t)$ for the same time. 
		Environmental fluctuations interfere with extinction process in a nonlinear way. Strikingly,  weak fluctuations decrease genetic loss and strong ones the intensify extinction process. (b) Heterozygosity as a function of  adimensional  time for  different values of the environmental noise amplitude $\xi$. Diffusion constant fluctuations postpone genetic loss in the range $0 < \xi < 0.2$. The inset shows that for  $\xi >0.2$  diffusion constant  fluctuations intensify genetic loss.}
	\label{FIG3}
\end{figure}

Number fluctuations are motivated from properties of the populations, here we find that fluctuations originated from the environment can also play a substantial role in genetic loss processes. The possible effect of the environment has been overlooked in many fields, including tumor genetics \cite{williams2016identification, mcdonald2018currently}. Our findings suggest that even simple properties of the environment can affect the frequency of competing clonal subpopulations and thus evolutionary dynamics within tumors.

These findings  may have implications in the context of human populations where different parameters such as the demographic modalities in which migration takes place, are known to drive expansion \cite{cavalli1993demic}. As such, spatial fluctuations in any of these properties can interfere with genetic loss and increase/decrease its intensity. This could explain in a simple way  how genetic loss can be faster in specific spatial areas \cite{keinan2007measurement}, or be faster than expectations for other cases \cite{keinan2009accelerated}. Physical models have already proven to be useful in prehistoric human populations studies by finding the velocity of  Neolithic transition (the shift from hunter-gatherer to agricultural economies) in Europe \cite{fort1999time, vlad2002systematic,fort2007fronts}. The model studied here could prove useful in a different context related to the evolution of the human species.

\textit{\textbf{Conclusion}}.  In this paper, we have analyzed the effect of spatial fluctuations of the diffusion on different properties of invading populations. Similar to  population number  fluctuations, diffusion fluctuations lead to front position wanderings.
As to genetic loss, our results show that while weak fluctuations favor the coexistence of populations, strong fluctuations intensify the extinction process. These findings together shed light on invasion in heterogeneous environments and provide grounds for the explanation of variations in human genetic loss intensity in different geographical regions  and provide another source of heterogeneity in human tumors evolutionary dynamics.

\textit{\textbf{Acknowledgments}.} We would like to thank G. F. Calvo (UCLM) for a critical reading of the manuscript. This work has been partially supported by Ministerio de Ciencia y Universidades/FEDER, Spain (grant MTM2015-71200-R), Junta de Comunidades de Castilla-La Mancha, Spain (grant SBPLY/17/180501/000154) and James S. Mc. Donnell Foundation (United States of America) 21st Century Science Initiative in Mathematical and Complex Systems Approaches for Brain Cancer (Collaborative award 220020560).

\end{document}